\documentclass[twocolumn,prd,floatfix]{revtex4}
\usepackage{color}
\usepackage{tabularx}
\usepackage{epsfig}
\usepackage{amsmath}
\usepackage{amssymb}
\usepackage{bm}
\usepackage{graphicx}
\usepackage{multirow}

\usepackage{epsfig}

\begin{document}

\title{Constraint on the mass of fuzzy dark matter from the rotation curve of the Milky Way}

\author{Alireza Maleki, Shant Baghram and Sohrab Rahvar}
\affiliation {Department of Physics, Sharif University of Technology, P. O. Box 11155-9161, Tehran, Iran.}

\date{\today}

\begin{abstract}
Fuzzy Dark Matter (FDM) is one of the recent models for dark matter. According to this model, dark matter is made of very light scalar particles with considerable quantum mechanical effects on the galactic scale, which solves many problems of the cold dark matter (CDM).   Here we use the observed data from the rotation curve of the Milky Way (MW) Galaxy to compare the results from FDM and CDM models. We show  FDM  adds a local peak on the rotation curve close to the center of the bulge, where its position and amplitude depend on the mass of FDM particles. By fitting the observed rotation curve with our expectation from FDM, we find that the mass of FDM is $m = 2.5^{+3.6}_{-2.0} \times10^{-21}$eV.  We note that the local peak of the rotation curve in MW can also be explained in the CDM model with an extra inner bulge model for the MW Galaxy. We conclude that the FDM model explains this peak without a need for extra structure for the bulge.
\end{abstract}

\pacs{}
\maketitle
\section{Introduction}
The rotation curves of the galaxies have been a useful tool for studying their kinematics and mass distribution \citep{oort1927observational,lindblad1927state,babcock1939rotation,deSwart:2017heh}. Based on these studies, it has been observed that the rotation curve outside a galaxy, reaches  a constant value, which is in contrast to our expectation of having a decrease in velocity according to  Kepler's law   \citep{freeman1970disks,rubin1970rotation}. This flat behavior of rotational velocity is very strong evidence in favor of an inconsistency existing between the theory and observation of galactic dynamics \citep{zwicky1933rotverschiebung,kahn1959intergalactic,page1959masses}. It is supposed that there must be a component for the structure of the galaxy, which is called the dark matter halo in literature, with a mass density that could provide a flat rotation curve for the galaxies. Dark matter is also needed on the cosmological scales to have the consistency with dynamics and structure formation based on observations \citep{peebles1993principles,bertone2005particle}.

However,  there are modified gravity theories trying to explain the dynamics of galaxies without a need for the mysterious dark matter fluid \cite{milgrom1983modification,moffat2006scalar,rahvar1,rahvar2}. In the context of dark matter theories, understanding the properties and the nature of dark matter is one of the most challenging and active fields of study in physics and cosmology. Based on the standard cosmological constant-cold dark matter ($\Lambda$CDM) model, and the latest data acquired by the {\it{Planck}} satellite, the Universe consists of  $68\%$ dark energy and $28\%$ dark matter with $4\%$  baryonic matter \citep{aghanim2018planck}. The $\Lambda$CDM model considers nonrelativistic, collisionless particles with  very weak interaction with the baryonic matter \citep{bernabei2003dark,markevitch2004direct}. Simulations based on the $\Lambda$CDM model showed that it is very successful at explaining the Universe on large scales \citep{springel2005simulations}. However, it seems that the interpretation of the observations on small scales faces some serious problems \citep{weinberg2015cold}. The important problems are the "core-cusp" problem  \citep{alvarado1993observational,moore1999cold}, the "missing satellites" problem \citep{klypin1999missing}, and the "too big to fail" problem \citep{boylan2011too}.  We also should note that all efforts for the detection of dark matter particles have been unsuccessful \citep{aprile2017first}, although there are some observational ideas to detect dark matter subhalos on a galactic scale \cite{Baghram:2011is,Erickcek:2010fc,Rahvar:2013xya,Asadi:2017ddk}.

There have been many proposals to solve the small-scale problems of CDM  \citep{weinberg2015cold,hu2000fuzzy}, such as warm dark matter (WDM) \citep{avila2001formation}, self-interacting dark matter (SIDM) \citep{rocha2013cosmological}, feedback effects the baryonic matter on the profile of the halo \citep{governato2010bulgeless,garrison2013can}, or modified initial conditions \citep{nakama2017stochastic,kameli2019modified}.  Another interesting approach to solving these problems is to consider dark matter made of very light particles with a mass of ($m\simeq 10^{-23}-10^{-21}$ eV) with a large de Broglie wavelength, where their quantum properties on small scales (kpc) play an important role and enable this model to solve the small-scale problems of CDM \citep{hu2000fuzzy}. These particles form a Bose-Einstein condensation (BEC) in the halo's central region and result in a condensate core which, due to its solitonic properties, is often called a soliton \citep{bohmer2007can,schive2014understanding,Li:2020qva}. In this model, the quantum pressure stabilizes the gravitational collapse and prevents the formation of a cusp by suppressing the small-scale structures \citep{hu2000fuzzy,woo2009high,lee2010minimum}. Simulations based on this model also show that its prediction on a large scale is the same as that for the CDM, but at the small scales, it has different predictions, which is interestingly consistent with the observational data \citep{schive2014cosmic}. The observational tests, such as weak lensing of the colliding FDM cores and x-ray emission from the colliding area to test the hypothesis of FDM, are studied in Ref.\cite{maleki}. Also, in Ref.\cite{kendall2019core}, using the rotation curves of galaxies in SPARC data, a mass for the FDM is obtained as $m \geq10^{-23}$eV. Also, there have been studies on some tensions of the FDM model with observational data from the rotation curves of galaxies, which suggest that the study of the MW Galaxy could probe the FDM model particle mass of $m <10^{-19}$eV  \cite{bar2018galactic}.

In this model, the dark matter particles are assumed to be non-self-interacting scalar fields. One important candidate can be axionlike particles \citep{schive2014cosmic}. These particles are predicted in high-energy physics theories. For example, in the string theory, all models have at least several bosonic axionlike fields \citep{hui2017ultralight}. There are many names for this model of dark matter in the literature, such as "wave dark matter", "ultralight axionic particles (ULP)", "Bose-Einstein condensate dark matter (BEC DM)", "fuzzy dark matter (FDM)" \citep{lee2018brief}. Here we use the term "FDM" for this model. One main concern about the FDM model is to use observational data to investigate the existence of these types of particles as a candidate for dark matter. Our attempt is to introduce the observational features of FDM in the rotation curve of the MW Galaxy.
In Sec. (\ref{S2}) we discuss the FDM properties. In Sec.(\ref{S3}) we introduce a model for the MW Galaxy mass components and the contribution of each component in the rotation curve of the MW. We also calculate the rotation curve based on the FDM halo model for the MW and find the best value for the mass of the scalar field.  In Sec. (\ref{S4}), we compare the differences between the CDM and FDM models. Section \ref{conclusion} is the summary and conclusion of this work.
\label{key}


\section{ FDM halo core (soliton) }
\label{S2}
\begin{figure}
\centering
\includegraphics[scale=0.5]{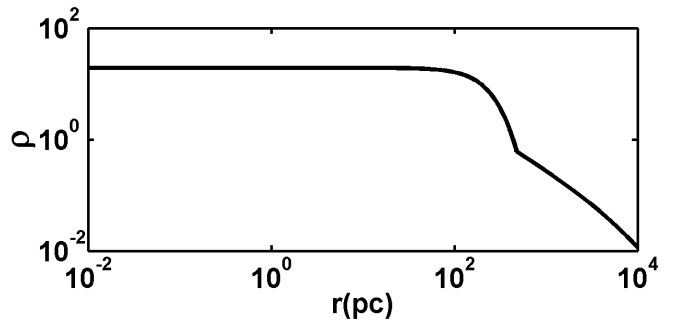}\caption{The mass density profile for the FDM halo model of the MW. We consider the FDM particles with $m=8.1 \times10^{-23}\text{eV}$. We use the solitonic mass density profile which has a core with the size of $r_c\sim200 \text{pc}$, the transition radius is $r_t\sim500 \text{pc}$. For $r>r_t$, the profile follows NFW. The x axis has  units of $\text{pc}$, and the density is plotted in  units of $M_{\odot}{\text{pc}}^{-3}$}
\label{solitonnfw}
\end{figure}

In this section, we will investigate the properties of a FDM halo core. We use Schrodinger-Poisson (SP) equations for the dynamics of a quantum system under its self-gravity \citep{paredes2018nonlinear,navarrete2017spatial} as
\begin{align}
i\hbar \frac{\partial \psi}{\partial t} =&-\frac{\hbar^{2}}{2m} \nabla^2\psi+mU\psi, \\
\nabla^2 U=&4\pi G\rho,
\label{psequation}
\end{align}
where $m$ is the mass of the FDM particles. The mass density is defined as $\rho={\mid \psi \mid}^{2}$ and $U$ is the gravitational potential. Simulations based on fuzzy dark matter with SP equations in a comoving frame show that the system will form a halo with a core. The profile of a FDM core, $\rho_{c}(r)$ which is called a soliton, is as below \citep{schive2014cosmic}:
\begin{align}
\rho_{c}(r)=\rho_{0}[1+0.091(\frac{r}{r_{c}})^2]^{-8},
\label{cdensity profile}
\end{align}
where $r_c$ is defined as the core radius and at $ r=r_{c}$ the density reduces to half of its peak value. The central mass density of the core is given by
\begin{align}
\rho_{0}=1.9 \times(\frac{10^{-23}\text{eV}}{m})^{2}(\frac{\text{kpc}}{r_{c}})^{4} M_{\odot}\text{pc}^{-3}.
\label{cdensity0}
\end{align}
Schive ~ et al. \cite{schive2014cosmic} found a relation between the halo mass $M_h$ and the core radius which is an empirical result from simulations
\begin{align}
r_{c}=1.6\text{kpc} \times(\frac{10^{-23}\text{eV}}{m})(\frac{M_h}{M_{\odot}\times 10^9})^{-1/3}.
\label{rcMh}
\end{align}
The mass enclosed in this radius is called the core mass, $M_c$. The outer region of the FDM halo behaves like an ordinary cold dark matter, which is well approximated by the  Navarro, Frenk, White (NFW) profile \citep{schive2014cosmic,du2016substructure}.  Generally, the full mass density of the halo can be written as
 \begin{align}
 \rho =\rho_{c}\theta(r_{t}-r)+\rho_{NFW}\theta(r-r_t),
 \label{fullFDMdensity}
 \end{align}
where $\theta$ is a step function and $r_t$ is  the scale on which the transition from BEC to ordinary phase happens. This specific scale is proportional to the core size (i.e., $r_t=\alpha r_c$), where $\alpha\sim 2-4$ \citep{schwabe2016simulations,mocz2017galaxy,robles2018scalar}.
Figure \eqref{solitonnfw} shows the mass density profile of the FDM halo for a MW-like galaxy.
For the outer regions, we use data for the rotation curve and the best value for the NFW parameters \cite{sofue2017rotation}. Also, if we adapt the mass of FDM as $m=8.1 \times10^{-23}$eV \citep{schive2014cosmic}, from Eq.(\ref{rcMh}), the core size would be $r_c\sim200\text{pc}$.
We also investigate the intersection of the core and NFW halo in the FDM model, using the continuity condition. With the adapted values for the mass of FDM, the transition radius for the MW Galaxy is $r_t\sim500\text{pc}$ (as shown in Figure \ref{solitonnfw}), which happens for $\alpha\simeq 2.5$. By decreasing the mass of FDM the core size increases, and according to Eqs (\ref{cdensity0}) and (\ref{rcMh}), the central density decreases. To satisfy the continuity condition between the density of NFW and the solitonic core, the mass of the FDM has to be larger than $m>10^{-23}$eV. In the next section, we will investigate the best mass for FDM particles to fit with the rotation curve of the MW Galaxy.

\section{The mass model for the Milky Way}
 \label{S3}
 In this direction, we model the mass distribution in the MW Galaxy by considering a bulge, a disk, and the dark matter halo, with a supermassive black hole at the center of the MW, which are the main mass components. It is worth mentioning that in reality, the system is more complex, containing the arms of the Galaxy and bars that cause asymmetric effects in the rotation curve,  but these features can be considered as second-order terms without considerable effects \citep{sofue2017rotation,sofue2013rotation}. So, we consider the main components of the MW with an additional solitonic structure at the center of the bulge.
 The circular rotation velocity of a test particle around a mass density is given by $$v^{2}(r)=Gr\int \frac{\rho(r')d^3r'}{|r-r'|^2}.$$
Accordingly, the total rotational velocity is a sum of all the components of the Galaxy as follows:
\begin{equation}
v_{rot}^{2}=v_{bh}^{2}+v_{b}^{2}+v_{d}^{2}+v_{dm}^{2},
\label{vroTotal}
\end{equation}
the subscripts "bh","b","d","dm" indicate the central back-hole, bulge, disk, and dark matter halo, correspondingly.
In what follows, we explain the model that we use for each component of the Galaxy.
\subsection{The bulge}
The mass profile for the bulge of a galaxy is parameterized by de Vaucouleur's suggestion \citep{de1958photoelectric,sofue2009unified}. It has been found through investigation that this profile cannot explain the MW's rotation curve in the central region. The suggested profile is made of two components: the inner bulge, and the outer bulge where the overall density profile can be written as
   \begin{equation}
  \rho_{b}(r)=\rho_{ibc} e^{-\frac{r}{a_{ib}}} + \rho_{obc}e^{-\frac{r}{a_{ob}}},
  \end{equation}
where $\rho_{ibc}$ and $\rho_{obc}$ are the central densities of the inner and outer bulge profiles,
and $a_{ib}$ and $a_{ob}$ are the corresponding characteristic core scales. Using this profile, the rotational velocity is
\begin{equation}
v_{b}^2(r)= \frac{GM_{ib}}{a_{ib}}\frac{F(x)}{x} +  \frac{GM_{ob}}{a_{ob}}\frac{F(y)}{y},
\label{vb}
\end{equation}
where we define  $x={r}/{a_{ib}}$,  $y={r}/{a_{ob}}$, $M_{ib}=8\pi a_{ib}^3\rho_{ibc}$, and $M_{ob}=8\pi a_{ob}^3\rho_{obc}$
are the total masses for the inner and the outer bulges. The function $F(z)$ is defined as
 \begin{equation}
F(z)=1- e^{-z}(1+z+\frac{z^2}{2}).
\end{equation}
\begin{figure}
	\centering
	\includegraphics[scale=0.60]{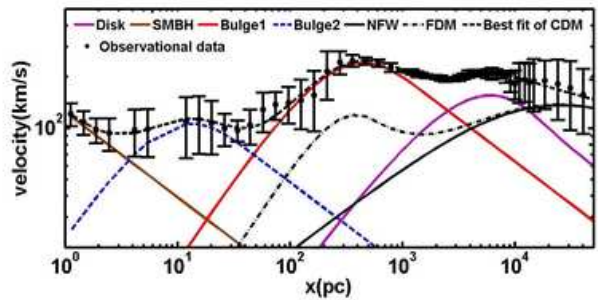}
	\caption{ The theoretical rotation curve of the MW from the NFW and FDM models. The overall rotation curve of the CDM model (dashed black line) results from the contributions of the central black hole (solid brown line), disk (solid pink line), bulge 1 (solid red line), bulge 2 (dashed blue line), and NFW halo (solid black line). The FDM rotation curve is plotted with the dash-dotted black line. Here we adapt the mass of FDM to be $m=8.1 \times10^{-23}$ eV \citep{schive2014cosmic}.The parameters of CDM and FDM are given in Table \ref{tab}.}
	\label{MWrcComponents}
\end{figure}
Figure \ref{MWrcComponents} shows the rotation curve of the MW \citep{sofue2013rotation,sofue2017rotation}, where on one hand, we will use the two-bulge model to fit the observational data, and on the other hand, we will apply the FDM model to explain the extra peak in the rotation curve of the Galaxy without needing to use the second bulge model.
\subsection{The disk}
The MW Galaxy has a baryonic matter disk which can be approximated by an exponential function.  The surface mass density of the disk \citep{mo2010galaxy} in cylindrical coordinates is given by
\begin{equation}
\Sigma_d(R)=\Sigma_0 e^{-\frac{R}{a_d}},
\end{equation}
in which  $\Sigma_0$ is the central value of the profile, and $a_d$ is the characteristic radius of the disk. The rotational velocity corresponding to the mass profile is
\begin{equation}
v_{d}(R)= \sqrt{\frac{GM_d}{a_d}}D(y),
\label{vd}
\end{equation}
where  $y={r}/{a_d}$, and $M_d=2\pi\Sigma_0 a_d^2$ is the total mass of the disk and the function $D(y)$ is defined as
\begin{equation}
D(y)=(\frac{y}{\sqrt{2}})\sqrt{I_0(\frac{y}{2})k_0(\frac{y}{2})-I_1(\frac{y}{2})k_1(\frac{y}{2})}
\end{equation}
The functions $I_i$ and $k_i$ are the first and second kinds of modified Bessel functions, respectively \citep{BT}.
  \subsection{Central supermassive black hole}
It is believed that there is a supermassive black hole at the center of the MW, known as Sagittarius $A^{\star}$, with a mass of $M\sim 4 \times10^6 M_{\odot}$ \citep{ghez1998high,narayan1995explaining,broderick2011evidence}. This black hole dominates the dynamics of the rotation curve in the inner regions and causes an increase of the rotation curve in the center of the MW Galaxy.
\subsection{Dark matter}
   \subsubsection{Cold dark matter}
    Dark matter is the dominant mass component in the Universe. Even though in the MW, the central region is mainly composed of baryonic matter, in the outer region the rotation curve is mostly obtained from the dark matter mass density. For cold dark matter, the well-known mass density profile is the NFW profile, which is \citep{navarro1995simulations}
   \begin{align}
    \rho(r)=\frac{\rho_ H}{R(1+R)^2}
    , \label{NFWP}
    \end{align}
   where $\rho_H$ is the characteristic density and $R$ is defined as $R={r}/{R_H}$, in which $R_H$ is the characteristic radius of the halo. So, the enclosed mass within the radius $r$ is
 \begin{align}
 M_h(r)=4\pi\rho_H R_H^3[\ln(1+R)-\frac{R}{1+R}],
 \label{NFWm}
 \end{align}
where the rotational velocity is obtained from $v_h^{2}(r)={GM_h(r)}/{r}$.
\subsubsection{Fuzzy dark matter}
The FDM mass density profile consists of a core and a transition to the NFW profile. Inside the core, using Eq.(\ref{cdensity profile}), the rotational velocity is
 \begin{equation}
 v_{rot}^{2}=\frac{4 \pi G \rho_{0}}{r} I_{1},
 \label{vrootpo}
 \end{equation}
where
 \begin{equation}
 I_1=\sum _{n=0}^{6}\frac{a_{n}r_c^2 r(\frac{r}{r_c})^{2n}}{(1+.091(\frac{r}{r_c})^2)^7}+a_7 r_c^3\arctan (a_8\frac{r}{r_c}),
 \label{point}
 \end{equation}
 in which the parameters are $a_0 = -0.1771$, $ a_1 = 0.2259$, $a_2 =  0.3907$, $ a_3 = 0.0030$, $ a_4 =0.0002$, $a_5 =  7.3664\times 10^{-6}$, $a_6 = 1.0055\times 10^{-7}$, $a_7 = 0.5870$, and $a_8 = 0.3017$. For the outer region, we use the Eq.(\ref{fullFDMdensity}), which considers the
 ordinary phase of FDM.
One important concern is the interaction of the supermassive black hole with the solitonic core. The effect of a supermassive black hole on the background scalar field of dark matter has been studied in the literature\citep{cardoso2018constraining,2017Barranco,bar2019looking,davies2020fuzzy}. Throughout this interaction, the soliton would be accreted by the black hole in timescales which depend on the black hole mass and  the mass of the FDM particles \citep{2017Barranco,davies2020fuzzy}. Accordingly, as discussed by Davies et.al (\citep{davies2020fuzzy}),  for the MW Galaxy in the case of $m\geq 10^{-19.4}$ eV, the soliton would be accreted by the black hole due to superradiance in a half-lifetime of less than the Universe's age. As we will see, this limit is larger than our results for the FDM particles' mass.
\section{comparison between the FDM and NFW cold dark matter models}
 \label{S4}
\begin{figure}
	\centering
	\includegraphics[scale=0.60]{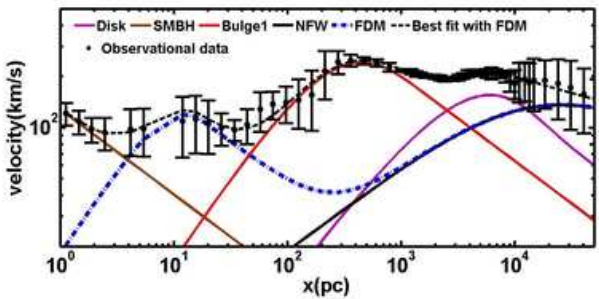}\label{diffMaFDM}
	\caption{ Rotation curve of the Milky Way. The dashed line is the theoretical rotation curve of the MW Galaxy in FDM with the best fit to the observed data. From the best fit we find $m = 2.5^{+3.6}_{-2.0} \times10^{-21}$eV . The inner bump of the MW rotation curve can be explained with FDM without needing the second bulge as used in the NFW model. The best values of the parameters of the FDM model are given in Table (\ref{tab2}).}
\end{figure}

\begin{figure}
	\centering
     \includegraphics[scale=0.60]{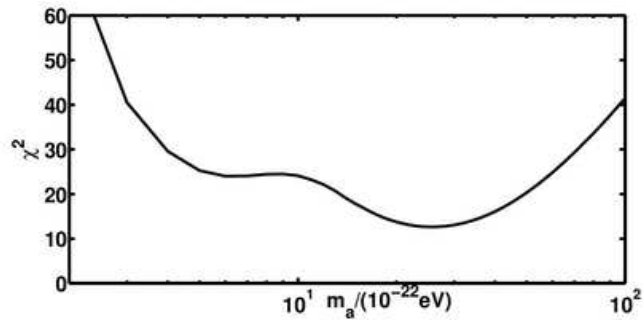}
	\caption{$\chi^2$ parameter  as  a function of the mass of FDM particles from fitting to the rotation curve of the Galaxy. The best value of this mass is $m = 2.5^{+3.6}_{-2.0} \times10^{-21}$eV.}
	\label{X2ma1}
\end{figure}

\begin{table}
\centering
\caption{The best values of the NFW model from fitting with the rotation curve of the MW Galaxy. The dashed black line in Fig. (\ref{S2}) represents the best fit to the rotation curve with a normalized $\chi_N^2$ of $0.0209$. The last row is not entered in the fitting process to the rotation curve. We use this parameter to calculate the
rotation curve from the solitonic part of the FDM structure in Fig.(\ref{S2}).}
\label{tab}
\begin{small}
\begin{tabular}{ | m{7em} | m{7em} | m{2cm}| }
\hline
Mass component&Mass$~(M_{\odot})$ &Characteristic length scale (kpc) \\
\hline
Supermassive black hole&$3.6^{+2.8}_{-2.0}\times10^6$& \\
\hline
Inner bulge& $5.4^{+5.0}_{-3.8}\times10^7$&$0.0040^{+0.0094}_{-0.0018}$ \\
\hline
Main bulge& $9.4{\pm0.6} \times10^9$& $0.134^{+0.029}_{-0.026}$ \\
\hline
Disk &$4.1\pm0.5\times10^{10}$ & $2.830^{+0.230}_{-0.190} $ \\
\hline
Dark matter (NFW)&$8{\pm2}\times10^{11}(r<1{\text{Mpc}})$&$12.0\pm2 $\\
\hline
Solitonic core ($8.1 \times10^{-23}$eV)&$M_{c}=4.1\times10^{9}$& 0.2\\
\hline
\end{tabular}
\end{small}
\end{table}%
In this section, we compare the two models of standard NFW dark matter and FDM for the rotation curve of the MW. Figure (\ref{MWrcComponents}) shows the rotation curves resulting from the cold dark matter halo and that of FDM. Here we adapt the mass of FDM particles to be $m=8.1 \times10^{-23}$ eV \citep{schive2014cosmic}. The rotation curve is not consistent with the CDM model at the inner bulge of the MW, and we need to add a second bulge with a mass of $5.4\times 10^7~M_\odot$ with the characteristic size of $4.1$ pc. This structure results in an additional contribution to the rotation curve of the central part of the Galaxy as indicated by the dashed blue line in Fig.(\ref{MWrcComponents}). The dashed black line in this figure represents the rotation curve from the NFW model. In order to fit the observational data, we allow the parameter of CDM to be free, and taking into account the second bulge structure, we find the best parameters as indicated in Table (\ref{tab}). Our results for NFW parameters are almost consistent with the values in Refs. \citep{sofue2013rotation,sofue2017rotation}.
In what follows we let the mass of FDM be a variable parameter whereby increasing the mass of FDM, the rotation curve in Fig. (\ref{MWrcComponents}) shifts to the left side of the diagram. Due to a natural bump in the rotation curve of FDM, we examine the possibility of replacing the contribution of the second bulge with the contribution of FDM in the rotation curve. We again allow the parameters of the CDM and FDM to be free parameters and find the best values for these two models. The results are given in Table (\ref{tab2}). Figure (\ref{diffMaFDM}) shows the rotation curve for the best values of FDM. Figure (\ref{X2ma1}) represents the dependence of $\chi^2$ on the mass of FDM. Here we find the best value for the mass of FDM to be $m = 2.5^{+3.6}_{-2.0} \times10^{-21}$eV with a corresponding $1\sigma$ error. With this mass for the FDM, we can fit the rotation curve of the Galaxy close to the Galactic core without needing to take into account the extra component to the bulge structure.
\begin{table}
\centering
\caption{The best values of the FDM model from fitting with the rotation curve of the MW Galaxy. The dashed line in Fig. (\ref{S3}) represents the best fit to the rotation curve with normalized $\chi_N^2 = 0.0214$. }
\label{tab2}
\begin{small}
\begin{tabular}{ | m{7em} | m{7em} | m{2cm}| }
\hline
Mass component&Mass$(M_{\odot})$ &Characteristic length scale(kpc) \\
\hline
Supermassive black hole&$3.8^{+2.7}_{-2.1}\times10^6$& \\
\hline
Main bulge& $9.4{\pm0.6} \times10^9$& $0.133^{+0.030}_{-0.026}$ \\
\hline
Disk &$4.1\pm0.5\times10^{10}$ & $2.830^{+0.230}_{-0.190} $ \\
\hline
Dark matter (NFW)&$8{\pm2}\times10^{11}(r<1 {\text{Mpc}})$&$12.0\pm2 $\\
\hline
Solitonic core ($2.5^{+3.6}_{-2.0}\times10^{-21}$eV)&$M_{c}=1.2^{+1.7}_{-1.0}\times10^{10}$&$0.006^{+0.002}_{-0.004}$ \\
\hline
\end{tabular}
\end{small}
\end{table}%
\section{summary and conclusion}
\label{conclusion}
In this work, we used the MW's rotation curve to examine FDM with the observational data. Also,
we compared it with the conventional Galactic model with the structures of one disk, two bulges (inner and outer), a central black hole, and a halo with the NFW profile. The inner bulge is needed in this model to generate the central bump in the rotation curve of the MW. We have shown that using  FDM as the dark halo structure not only produces the correct rotation curve at the outer parts of the Galaxy, but also plays the role of the central bulge to produce the same bump profile in the rotation curve of the Galaxy.
From the best fit to the rotation curve of the MW, we find the parameters of luminous components and the halo in both NFW and FDM halo models. From the best fit, we obtain the mass of the FDM to be $m = 2.5^{+3.6}_{-2.0} \times10^{-21}$eV with a core mass of $M_c = 1.2\times 10^{10} M_\odot$. Using the continuity condition for the mass profile of the core and the outer halo where dark matter transits from the BEC at the central part of the halo to the NFW state puts a lower bound of $m \geq10^{-23}$ eV on the mass of FDM particles. \\ \\
\section*{Acknowledgments}
 S.B. is partially supported by the Abdus Salam International Center of Theoretical Physics (ICTP) under the junior associateship  scheme during this work. This research is supported by the Sharif University of Technology Office of Vice President for Research under Grant No. G960202. \\ \\

\end{document}